# An Extra Dimensional Approach of Entanglement

Axel Dietrich[1]*  &  Willem Been[2]

[1] )Institute of Human Genetics, [2] ) Department of Anatomy and Embryology,
*University of Amsterdam., AMC M-1, Meibergdreef 15, NL 1105 AZ Amsterdam., The Netherlands*
* Corresponding author, e-mail: a.dietrich@amc.uva.nl

Motivated by the apparent lack of a workable hypothesis we developed a model to describe phenomena such as entanglement and the EPR-paradox.
In the model we propose the existence of extra hidden dimensions. Through these dimensions it will be possible for particles, which originate from one source, to remain connected. This connection results in an instantaneous reaction of one particle when the other particle is manipulated. We imagine entanglement in such a model. The results of the experiments which have been performed on this item do not contradict with the existence of the extra dimension(s).
In addition, the model opens the possibility to unify the theory of quantum mechanics, gravitation and the general theory of relativity.

### Introduction
In 1935 Einstein, Podolsky and Rosen initiated a discussion that continues up until now and which forms the basis for many experiments [1]. They put forward serious criticism on the validity of the quantum theory. They stated in their paper that the quantum mechanical description of reality is not complete. Or, when operators corresponding to two physical quantities do not commute the two quantities cannot have simultaneous reality. They developed a thought experiment in which two systems interact for a short period of time, after which there was no longer any interaction. It would be impossible to determine the exact properties of simultaneously appearing realities.
Bohm [2] expanded the formulation of the thought experiment using a particular example of two particles that originated from one source with opposite spin angular momentum. In that case he was unable to prove that the world is actually made up of separately existing and precisely defined elements of reality. To solve this problem Bohm introduced hidden variables. However, the quantum theory is inconsistent with the assumption of hidden causal variables. As a consequence, this was in agreement with Einstein, Podolsky and Rosen's claim that the theory of quantum mechanics is not complete enough to describe and predict these phenomena [1].
Bell [3] formulated an inequality principle to test for the existence of hidden variables. He showed that there could be no local hidden variables if the inequality was satisfied.

### Experiments
Alain Aspect and co-workers [4, 5] performed excellent experiments to determine the mutual influence of a pair of photons which originated from one single source. The correlation of linear polarizations of pairs of photons was measured. It was demonstrated that the obtained results were in agreement with the quantum mechanical predictions. This was a strong violation of the generalised Bell's inequalities. The effect which was measured was carried out over 13 m in a time of 10 nanoseconds [6]. To bridge a distance of that size with the speed of light about four times of the observed time is necessary. This excludes an interaction in which elements are exchanged with the speed of light.
The possibility of selection cannot be excluded in the experiments of Aspect and his group. Because they use polarimeters there is a selection on a certain type of photons and the polarization states are determined. In addition, Aspect et al. [5] used periodic sinusoidal switching, which is predictable into the future. This does not exclude an explanation by communication slower than the speed of light [7]. Aspect and his group did not perform actually manipulation on the photon after which the state is determined of its twin photon that originated from the same source. The polarization states could actually be the initial states that have existed from the beginning. The group of Anton Zeilinger [7] claims to be the first who actually change the polarization state of a photon and determine this state of the corresponding photon over a distance of 400 m. During this period the quantum mechanical correlation is conserved. This is the correlation we know as entanglement.

**Description**
The description of the correlation of two particles, which in most cases originates from one source is called entanglement.
All experiments suggest that there is more than just well preserved initial properties of the two particles. The explanation nowadays is described by the term entanglement. Schrödinger [8] was the first to describe the phenomenon of a more than classical correlation of two particles, originating from one source. He introduced the term: "Verschränkung" which is entanglement as we call it nowadays.
Shimony [6] describes an additional correlation in the route the entangled particles follow. They represent a mirror image of the pathways which is followed. There must be a certain type of connection as stated by Shimony as: "striking correlation in their behaviour, so that a measurement done on one of the entities seems instantaneously to affect the result of the measurement on the other" [6], or as put by Einstein: "spooky actions at distance" [9]. They all agree on the fact that there has to be some kind correlation.
Why do we want to produce a more or less classical-like explanation of a phenomenon as entanglement? Because entanglement in our vision is a description and it does not give an explanation in the classical way, probably because an explanation it is very difficult to give. We think that a model is necessary to link different mechanisms and make a comparison possible. So we will try to represent an image that might give a clue towards an explanation. In addition, we would like to produce a model in which quantum mechanical phenomena can be linked onto the other fields in physics, because there are particles that obey the laws of both quantum mechanics as well as relativity. For example photons behave according to the quantum description in all experiments on entanglement. However, they also follow the laws of general relativity and are influenced by gravity. We developed a way of vision to match both phenomena.

**Possible explanations for the phenomenon**
What are the possible explanations for entanglement?
We do not expect that entanglement is just a perfect conservation of the initial qualities. This could hardly explain choice of the pathway both particles take as described by Abner Shimony [6]. There are no workable models in the classical sense, which give an explanation for phenomena such as entanglement.
We can think of a number of possible explanations:

1) Over the given distance the results of the experiments, that are performed up until now, might give the impression of the existence of a messenger which moves faster than the speed of light. However, messengers using a speed exceeding the speed of light are forbidden in the theory of special relativity [10]. As a consequence, we will not take this option into consideration.

2) The inexplicable correlation of the particles could be the result of a field comparable with the electromagnetic field. We think this is unlikely, because the observed results cannot be realised by phenomena such as fields, because such a field would influence all particles with equal qualities and it would seem that all those particles are entangled.

3) As a last option we could think of a connection between the particles, or even that the particles remain a unity. In these cases, because we cannot see the connection, it probably has to be in an extra hidden dimension. We think that this is the only possible explanation.

**The extra dimension**
Since Kaluza and Klein in the early 1920s suggested a hidden fifth dimension there has been an ongoing search for extra dimensions, which recently gave rise to a discussion [11, 12, 13]. The dimensions which are described up until now are very small ($10^{-35}$ meter), also in the Calabi-Yau setting [12]. There is a need for bigger extra dimensions. There is an ongoing search for them. It would be a great satisfaction when an extra dimension of considerable size could be found. A start of the experimental search for large extra dimensions has been made [14]. Perhaps we do not have to continue this search, because the large dimensions are already found, as we try to demonstrate in this paper.
We try to solve EPR/ Bohm /Bell/ Aspect contradiction using extra and invisible dimensions as looked for by Arkani-Hamed et al. [11], Antoniadis et al. [13] and Greene [14] of a considerable larger size than 1 mm. In the case of the experiments of Aspect and co-workers [4, 5] more than 10 m was observed and up to 400 m was observed by Weihs et al. [7].

We think that it is plausible that we found strong indications for the existence of these dimensions indirectly, without realising that we did so. Because the main goal of this paper is the production of visual insight into a mechanism, we will not use mathematics to describe the observed or expected phenomena. We just try to find a model that explains several phenomena such as entanglement and the EPR-paradox. For this reason we refer to Ludwig Wittgenstein [15] who pointed out that the use of mathematics is a method to describe certain phenomena, but that it is not necessary for the explanation. We will try to approach the explanation of entanglement in a describing manner.

**How should we imagine this?**
The best way is to imagine the situation in 2-D, where the 2-D position corresponds with the actual 3-D situation, whilst the 3D figure actually corresponds with the extra dimensional position (figure 1). Imagine a single event, for example the decay of an atom or molecule producing two particles with different spin angular momentum, like Bohm suggested [2]. It is our suggestion that the two particles will remain connected through an extra invisible dimension. At the beginning there is one particle and it will remain a unity because there is a connection by means of a certain type of super string, or even the real particle structure, running through one of the extra invisible dimensions (figure 1). This causes the instantaneous so called mutual influence.

The dimension can be considered as described in figure 1a. However, as an alternative one could consider the three dimensions in which we live, to be folded as suggested Arkani-Hamed et al. [11] (see figure 1b). In essence there is no difference between figure 1a and figure 1b. In the case of figure 1b the distance of e.g. 400 m in the 3D-world could be less in the extra dimension.

For both cases the circular endings are the only parts that can be observed in our limited familiar three dimensional world. When we manipulate one end, the connected corresponding other end will react instantaneously. This can also explain the change of the route as described by Shimony [6]. This way it can be explained why it seems that there is a messenger system which exceeds the speed of light. It certainly can explain phenomena that correspond with the description of entanglement.

The connection through the extra dimension should not necessarily consist of one string, or the particle unity. It could be considered as composed of more components, such as a chain of closed superstrings.

In conclusion, there is no need for a messenger between the particles with a speed exceeding the speed of light. It is a matter of unity: There is no inequality of Bell because the local qualities of two particles are in essence a unity going through an extra timeless dimension.

a

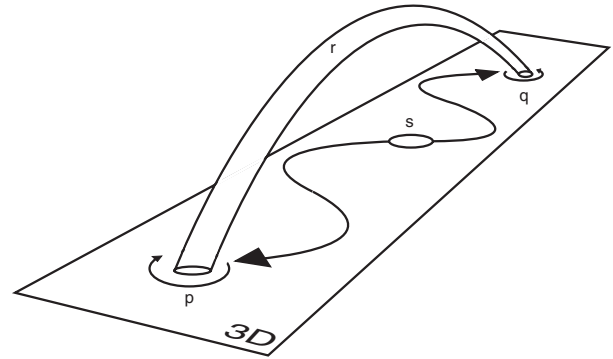

b

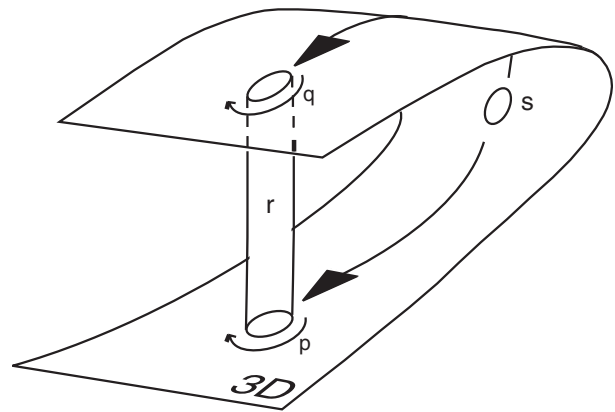

FIG. 1. Schematic representation of the connection through an extra dimension. a and b are alternative representations.
p and q = two particles of one pair originating from one initial event s. s = source of the particle pair. → and ← = quality such as the spin angular momentum of the different particles. r = extra dimension.

**Experimental indications**
The results of the experiments of Alain Aspect et al. [4, 5] and Weihs et al. [7] actually can be considered as a strong indication for the existence of the large extra invisible dimensions as searched for by Arkani-Hamed et al. [11]. In this case not for dimensions of 1mm but even up to more than 10 m. The experiments of Weihs et al. [7] demonstrate an entanglement even over a distance of 400 m after an actual manipulation of the quantum qualities of the particles. Nevertheless the size could be considerably smaller when the model presented in figure 1b is used. We could even consider the possibility that the distance approaches zero, in which case we observe one entity from two "sides".

Arkani-Hamed and his colleagues wonder why no larger extra dimensions are observed [11]. As a matter of fact quite larger extra dimensions in the results of the experiment of Alain Aspect and Anton Zeilinger and their groups can be seen. In our vision these results actually demonstrate the extra dimensions.

Our proposed model especially opens the opportunity to unify the theories of quantum mechanics, gravitation and general relativity [16]. This aspect of the model can be considered as the introduction of a broader unifying theory. We will try to present a more general theory in a separate paper.

Furthermore, the observed dimensions are rather large up to more than 400 m. It is not clear at what distances the elasticity of the dimensions or strings will remain intact or at what distance the elasticity will snap. In the case of the version of figure 1b this is not necessary.

It is unclear if the distance gets so large that the strings will disintegrate when they reach the end of elasticity, as proposed by Wheeler [17]. Aharonov [18] speaks of entanglement length and distinguishes finite and infinite entanglement length which depends on the background noise. Below the critical background noise the entanglement length is infinite [18].

When we consider the Big Bang as a starting point, or even before the Big Bang as suggested by Tryon [19], there might be found the initiation of gravity. He considered the gravitational energy as the opposite component of mass, giving a net energy of our universe of zero. Probably Tryon imagined that mass particles would remain connected by gravity. He did not mention extra dimensions but these could explain the mechanism. Particles could remain connected by gravitation through extra dimensions as imagined in figure 1a. In that case particles could remain connected, but perhaps the connections are disrupted at larger distances when the end of the elasticity is reached [17] or when the background noise becomes too strong [18].

**Perspectives**

It will be very useful to repeat the entanglement experiments, not by determination of polarization but by the study of the actual spin angular momentum and altering this of the particles as proposed by Bohm [2]. This is within the possibilities because there is system in which spin angular moment can be used on $^{199}$Hg as pointed out by Fry and Walther [20]. In this type of experiment undesired selection like polarization, can be avoided. In addition it probably will be useful to develop experiments in which other qualities than spin angular moment are used. This could even go as far as the annihilation of one of the two particles.

Further experiments will be necessary to determine whether there is "snapping" of strings at greater distance. Experiments as performed by the group of Anton Zeilinger [7] have to be carried out on a larger scale, for example in space, to determine what the maximum size of the extra dimension(s) is.

In addition, we think that our model will open the gate towards the possibility to unifying quantum mechanics and the theory of relativity.

**Acknowledgements**
We thank Ruud van den Bogaard for useful suggestions and discussion, Eelco Roos and Rob Lutgerhorst for producing the figures.